\documentclass[aps,prapplied,twocolumn,superscriptaddress,groupedaddress]{revtex4}

\usepackage{graphicx}% Include figure files
\usepackage{dcolumn}% Align table columns on decimal point
\usepackage{bm}% bold math
\usepackage{mathtools}
\usepackage[colorlinks=true]{hyperref}% add hypertext capabilities
\usepackage[dvipsnames]{xcolor}%citation colors
\usepackage[mathlines]{lineno}% Enable numbering of text and display math

\newcommand\myshade{85}
\colorlet{mylinkcolor}{Red}
\colorlet{mycitecolor}{Blue}
\colorlet{myurlcolor}{Gray}

\hypersetup{
  linkcolor  = mylinkcolor!\myshade!black,
  citecolor  = mycitecolor!\myshade!black,
  urlcolor   = myurlcolor!\myshade!black,
  colorlinks = true,
}

\newlength{\thelinewidth}
\thelinewidth=\textwidth
\begin{document}

\title{Phase-modulated cavity magnon polaritons as a precise magnetic field probe}

\author{Nicol\`o Crescini}
	\email{nicolo.crescini@phd.unipd.it}
	\affiliation{Dipartimento di Fisica e Astronomia, Via Marzolo 8, 35131 Padova, Italy}
	\affiliation{INFN-Laboratori Nazionali di Legnaro, Viale dell'Universit\`a 2, 35020 Legnaro (PD), Italy}
	
\author{Giovanni Carugno}
	\affiliation{Dipartimento di Fisica e Astronomia, Via Marzolo 8, 35131 Padova, Italy}
	\affiliation{INFN - Sezione di Padova, Via Marzolo 8, 35131 Padova, Italy}
	
\author{Giuseppe Ruoso}
	\affiliation{INFN-Laboratori Nazionali di Legnaro, Viale dell'Universit\`a 2, 35020 Legnaro (PD), Italy}
	
\date{\today}

\begin{abstract}
We describe and operate a spin-magnetometer based on the phase modulation of cavity magnon polaritons. 
In this scheme a rf magnetic field is detected through the sidebands it induces on a pump, and the experimental configuration allows for a negligible pump noise and a high frequency readout.
The demonstrator setup, based on a copper cavity coupled to an yttrium iron garnet sphere hybrid system, reached a sensitivity of $2.0\,\mathrm{pT/\sqrt{Hz}}$ at 220\,MHz, evading the pump noise and matching the theoretical previsions. An optimized setup can attain a rf magnetic field sensitivity of about $8\,\mathrm{fT/\sqrt{Hz}}$ at room temperature. An orders of magnitude improvement is expected at lower temperatures, making this instrument one of the few magnetometers accessing the sub-fT limit. 
%Due to its natural applications, miniaturization and multiplexing are eventually discussed.
\end{abstract}

\maketitle

\section{Introduction}
\label{sec:intro}
The detection of ultra-low magnetic fields is a long standing technological challenge.
% motivated by its many applications.
The most sensitive types of magnetic field detectors are essentially two: superconducting quantum interference devices (SQUIDs) \cite{PhysRevLett.12.159,doi:10.1063/1.322574,Aprili2006,1335547,doi:10.1002/3527603646} and spin-exchange relaxation-free (SERFs) \cite{Kominis2003,PhysRevLett.95.063004,Budker2007} magnetometers, which reach the $\mathrm{fT}/\sqrt{\mathrm{Hz}}$ sensitivity. Medical applications, such as nuclear magnetic resonance (NMR) detection, is possibly the most studied utilization of such devices \cite{PhysRevApplied.13.051002}. Remarkable results have been obtained with both SQUIDs \cite{RevModPhys.70.175,doi:10.1146/annurev.bioeng.9.060906.152010,Zotev_2007,Fedele_2015,Storm_2016,4277368,Storm_2019} and SERFs \cite{PhysRevLett.94.123001,SAVUKOV2007214,doi:10.1063/1.2392722,PhysRevLett.104.133601,PhysRevLett.89.130801}.
In the last decades several magnetic field sensors were proposed or demonstrated \cite{doi:10.1063/1.4960171,Taylor2008,Luomahaara2014,PhysRevLett.116.240801,PhysRevLett.108.120801,PhysRevA.76.023407,PhysRevB.84.214502,PhysRevLett.116.190801,PhysRevA.100.023416,Kuwahata2020,PhysRevLett.116.190801,wolski2020dissipationbased,doi:10.1063/5.0024369}, making precision magnetometry an alive and dynamic field.
SQUIDs are based on Josephson junctions and  require cryogenic temperatures to be operated.
Room temperature devices  are advantageous for cost and possible applications. SERFs magnetometers do not need cryogenics, but require heavy shielding and their sensitivity is volume-dependent.

This letter presents a magnetometer which embodies room temperature operation, size-independent sensitivity, and rf working frequency. 
It is based on the phase modulation of a pump field exciting a cavity magnon polariton resonance. Measurements of extremely weak fields are possible by the detection of a sideband component induced on the input pump, allowing for the sensing of signals from a few MHz to some GHz.
Its expected  limit  sensitivity is competitive with present state-of-the-art magnetometers and could find application, for example, in medical imaging, communication, and fundamental physics research \cite{articlecomm,doi:10.1063/1.1136907,Sternickel_2006,Budker2007}.

The working principle of this device is based on the excitation of the Larmor transition in a magnetic material. By taking a ferrite or a paramagnet and biasing it with a static field $\mathbf{B}_0$ we obtain an electron spin resonance at the frequency $\omega_\mathrm{L}=\gamma B_0$, where $\gamma=(2\pi)28\,$GHz/T is the electron gyromagnetic ratio. The presence of a weak magnetic field $\mathbf{b}_1$, with frequency $\omega_b$, has different effects. If $\mathbf{b}_1$ is perpendicular to $\mathbf{B}_0$ and on-resonance with $\omega_\mathrm{L}$, it will accumulate power into the Larmor transition, whose amount is proportional to $\mathbf{b}_1$ and can be used to measure it \cite{PhysRevLett.124.171801,quaxepjc}. In the complementary case where the direction of $\mathbf{b}_1$ is parallel to the static field, the total field determining the Larmor transition is time modulated and will periodically change the frequency of the spin resonance. A microwave tone pumping  the magnetic material at the frequency $\omega_\mathrm{L}$ will thus be phase modulated due to $\mathbf{b}_1$ with the result of producing sidebands at frequencies $\omega_\mathrm{L}\pm n \omega_b$, with $n$ an integer number.
The detection of the sidebands will eventually give a measurement of the weak perturbing magnetic field \cite{doi:10.1063/5.0024369}. The device we are presenting in this letter originate by embedding the second scheme in a photon magnon hybrid system (PMHS). 

\section{Detection scheme}
\label{sec:detection}
A PMHS consists in two coupled harmonic oscillators, e.\,g. the normal mode of a microwave cavity and the electron spin resonance of a magnetic material \cite{Lachance_Quirion_2019,PhysRevLett.111.127003,Clerk2020}.
A typical realization of such a system is obtained by properly inserting in a copper microwave cavity a sphere of yttrium iron garnet (YIG), a ferrimagnetic insulator with high spin density and low losses \cite{CHEREPANOV199381,PhysRevLett.3.32,PhysRev.110.1311}.
Antennas coupled to the cavity are used for inserting and extracting power from the system.
An external static magnetic field $\mathbf{B}_0$ is used to tune the Larmor frequency of electrons in YIG to the cavity mode frequency and couple the two systems.
For $\omega_\mathrm{L}$  close to the cavity frequency $\omega_0$, the cavity mode splits into two hybrid modes with frequencies $\omega_1$ and $\omega_2$, where $g_{21}=\min(\omega_2-\omega_1)/2$ is the coupling strength. If $g_{21}>\gamma_{1,2}$, the hybrid modes linewidth, the system is in the strong coupling regime.
The resulting system can be described by quasiparticles called cavity magnon polaritons (CMPs), whose dispersion relation is an anticrossing curve \cite{Kittel2004,landau,walls2007quantum}. 

Let us consider an oscillating field $\mathbf{b}_1$ parallel to the static one,  and apply a monochromatic tone on one of the hybrid modes, like the one at the frequency $\omega_2$. As in the case of the simple system described before, the oscillating field will produce a frequency modulation of $\omega_2$, thus causing a phase modulation of the electromagnetic field stored in the PMHS, and sidebands are generated at the frequencies $\omega_2 \pm n \omega_b$. However, since the PMHS has only two resonant modes, for $\omega_b > \gamma_2$, only the sideband satisfying the relation $\omega_2 \pm n \omega_b = \omega_1$ will be present and all others will be suppressed. By detecting the power at $\omega_1$, it will be possible to obtain a measurement of the perturbing field. For small fields the only component with significant amplitude corresponds to $n=1$ and this system is sensitive to fields whose frequency is $\omega_b=2g_{21}$.
With a PMHS the transduction coefficient between $B_0$ and mode frequency is not given anymore by the gyromagnetic ratio but to a smaller coefficient $\alpha\gamma$. Such coefficient, extracted from  $\partial \omega_{1,2}/\partial B_0$, is practically $\gamma$ for hybrid modes far away from the cavity resonance, while goes to zero for an hybrid mode close to it. In the case of full hybridization, namely $\omega_2 - \omega_0 = \omega_0 - \omega_1 =g_{21}$, we have $\alpha = 1/2$.

	\begin{figure}%[h!]
	\includegraphics[width=.4\textwidth]{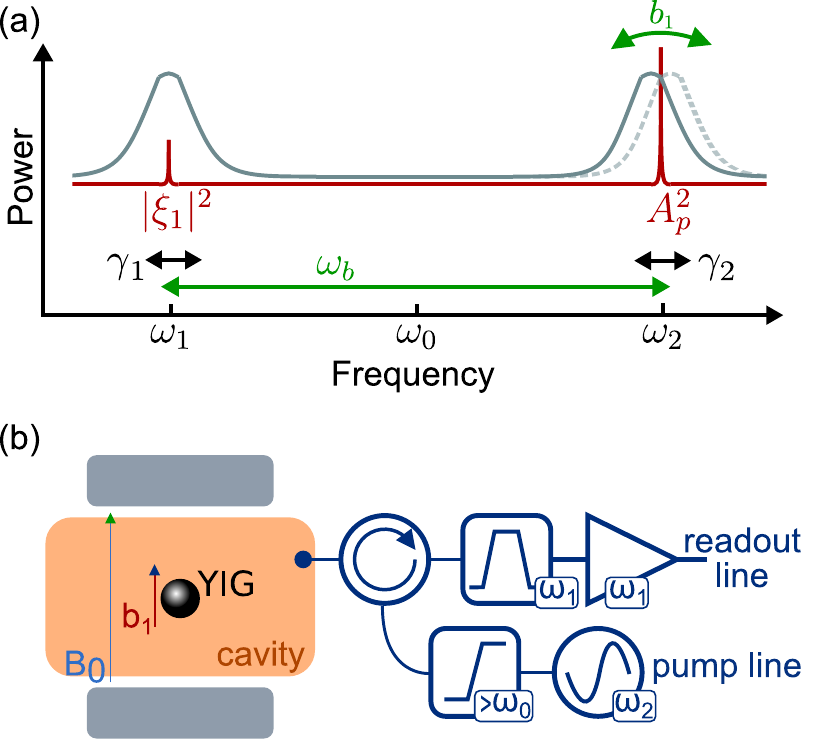}
	\caption{(a) Scheme of the spin-magnetometer working principle. The gray line represents the transmission spectrum of the PMHS, the dark red line is the pump with the corresponding sideband, the magnetic field effect and its frequency are reported in green. (b) Electronics of an example device. The pump line is filtered with a waveguide with cutoff around $\omega_0$ before exciting the $\omega_2$ resonance. An antenna (full dot) is used to inject the tone and extract the sideband. The output power is filtered and amplified before the readout.}
	\label{fig:scheme}
	\end{figure}

A scheme of the spin-magnetometer  principle is reported in Fig.\,\ref{fig:scheme}a. The monochromatic tone can be written as $A_0\cos(\omega_2t+\phi_b(t))$, where $\phi_b(t)$ is an oscillating phase added by the field $|\mathbf{b}_1|=b_1 \sin (\omega_b t)$.
The resulting phase modulated signal is
	\begin{equation}
	\xi(t)=A_0 e^{-i\omega_2 t} e^{-ix\sin(\omega_b t)},
	\label{eq:sig}
	\end{equation}
where $x$ is the  modulation index  given by the maximum deviation divided by the working frequency
	\begin{equation}
	x=\pi \alpha Q_2 b_1/ B_0.
	\label{eq:modindex}
	\end{equation}
Here, $Q_2=\omega_2/\gamma_2$ is the quality factor of the hybrid resonance.
For small values of $x$, Eq.\,(\ref{eq:sig}) can be recast in terms of the first kind Bessel functions $J_n(x)$ as
	\begin{align}
	\begin{split}
	\xi(t) = & \sum\limits_{n=0}^{+\infty}\xi_n(t)  = A_0 e^{-i\omega_2 t} \sum\limits_{n=0}^{+\infty}J_n(x)  e^{-i n \omega_b t} \\
	\simeq & A_0 e^{-i\omega_2 t}\sum\limits_{n=0}^{+\infty}\frac{x^n}{2^n n!}e^{-i n \omega_b t},
	\label{eq:sig_bessel}
	\end{split}
	\end{align}
where $n$ is a positive integer.
The amplitudes of carrier and lower order sidebands are 
%obtained by recasting Eq.\,(\ref{eq:sig_bessel}) e.\,g. for $n=0,1,2$ as
	\begin{equation}
	|\xi_0|= A_0,~ |\xi_1|=\frac{\pi \alpha A_0 Q_2 b_1}{2B_0}, ~ |\xi_2|= \frac{1}{8} \left( \frac{\pi \alpha A_0 Q_2 b_1}{B_0} \right)^2.
	\label{eq:sideband_amp}
	\end{equation}
The first order sideband $\xi_1$ has the higher amplitude, and by comparing it with the  backgrounds it allows to calculate the magnetometer's sensitivity to $b_1$.
%{\color{blue}It is clear from the above discussion that magnetic field sensitivity is only for ac field modulated at a frequency of order $2g_{21}$, in general in the range of tens to hundreds of MHz.}

The  scheme for a possible experimental set-up is shown in Fig.\,\ref{fig:scheme}b. A signal generator provides the tone at $\omega_2$,  filtered by a waveguide with cut-off frequency between $\omega_1$ and $\omega_2$. The waveguide reduces the pump noise at the frequency $\omega_1$  by a factor $k_1$ that could be as large as 80 dB, depending on the separation between the cut-off and pump frequencies. The waveguide output is fed to the PMHS, namely a resonant cavity with YIG inside, in the total field $\mathbf{B}_0+\mathbf{b}_1$. The antenna has couplings $\beta_{1,2}$ to the two resonant modes, respectively. The power reflected from the cavity is band-pass filtered and amplified before the readout. The filtering removes the residual pump tone  avoiding saturation of the  amplifier.
	
In an experimental setup we consider $Q_2=Q_2^{(0)}/(1+\beta_2)=Q_2^L$, the loaded quality factor of the CMP resonance.
The output power at the frequency $\omega_1 = \omega_2 - \omega_b$, normalized with the amplifier gain $G_1$ and filter transmittance $c_1$, results
\begin{widetext}
	\begin{align}
	\begin{split} 
	\frac{P_s (\omega_1)}{G_1 c_1} & = \frac{\beta_1}{1+\beta_1}  |\xi_1|^2  + k_1 \Big( \frac{1- \beta_1}{1+\beta_1} \Big)^2 A_p^2 |{\rm RAM}(\omega_b)|^2 \delta\nu+ k_B \left[ \frac{4\beta_1}{(1+\beta_1)^2} T + T_n\right]\delta \nu   \\
	& = A_p^2\left[\frac{\beta_1}{1+\beta_1}\frac{4\beta_2}{(1+\beta_2)^2} \left|\frac{\pi\alpha Q_2^{(0)} b_1}{2(1+\beta_2)B_0}\right|^2+ k_1 \Big( \frac{1- \beta_1}{1+\beta_1} \Big)^2 |{\rm RAM}(\omega_b)|^2\delta\nu\right] + k_B \left[\frac{4\beta_1 T}{(1+\beta_1)^2} + T_n\right] \delta \nu.
	\label{totpower}
	\end{split}
	\end{align}
\end{widetext}
Here ${\rm RAM}(\omega_b)$ is the relative residual amplitude modulation of the pump at $\omega_b$. The absorbed pump power at $\omega_2$ is $A_0^2=4\beta_2 A_p^2/(1+\beta_2)^2$, $k_1$ is the waveguide transmittance at $\omega_1$,  $k_B$ is the Boltzmann constant, $T$ the cavity temperature, $T_n$ the detection chain noise temperature and $\delta \nu$ the resolution bandwidth. Cables and circulators have been considered lossless.
The first term present in Eq.\,(\ref{totpower}) is the searched-for signal, and consists essentially in the sideband power $|\xi_1|^2$. The following terms are the sources of noise, which are related to the pump and to the system noise temperature. 
The typical behavior of ${\rm RAM}(\omega)$ shows large values at low frequencies with $1/f$-like trend up to a few MHz. 
All the terms are weighted with the appropriate combination of the antenna coupling constant to the first hybrid mode $\beta_1$.
At more than 10\,MHz the noise of a commercial pump is flat  with values $|{\rm RAM}(\omega)|^2\simeq -160$ \,dBc/Hz; the best oscillators reduce this limit down to -180\,dBc/Hz \cite{doi:10.1063/1.4726122}. This noise can then further reduced to a negligible level by using the waveguide filtering accounted by the term $k_1$, obtaining a thermal-noise limited device. 
In this limit we calculate the sensitivity $\sigma_{b_1}$ of the apparatus, defined as the value of $b_1$ giving a $\mathrm{SNR} = 1$ in the unit time,
\begin{equation}
\sigma_{b_1} = \frac{2\eta B_0}{\pi \alpha Q_2^L}\sqrt{\frac{k_B T_s}{A_p^2}},
\label{eq:sigmab1}
\end{equation}
where the antennas coupling correction $\eta$ and system noise temperature $T_s$ are
\begin{equation}
\eta = \frac{(1+\beta_2)}{2}\sqrt{\frac{1+\beta_1}{{\beta_2 \beta_1}}}, \quad T_s=\frac{4\beta_1T}{(1 +\beta_1)^2} + T_n.
\label{sens}
\end{equation}
With typical values $\beta_{1,2}\simeq 1$, $Q_2^L=10^4$, $B_0=0.4\,$T, $A_p^2 = 100\,$mW, $T_s= 350\,$K, and by working in the dispersive regime (i.\,e. with $\alpha\sim1$) a sensitivity $\sigma_{b_1} \simeq 8\,$ fT/$\sqrt{\rm Hz}$ is expected already at room temperature.

Notably, the sensitivity reported in Eq.\,(\ref{eq:sigmab1}) is unrelated to any extensive property of the setup. This is expected in a scheme based on phase-modulation rather than  absorption \cite{PhysRevLett.124.171801,doi:10.1063/5.0024369}, and paves the way to the miniaturization of the sensor without expecting a sensitivity loss. It needs to be considered that a strongly coupled PMHS often requires a large quantity of material, which might effectively limit the down-scaling of the setup. However, this can be compensated by a more concentrated cavity field, which increases the single spin coupling and allows for using less magnetic material.

This magnetometer only works in a narrow frequency band corresponding to the linewidth $\gamma_1$ of the detection mode $\omega_1$.
For a strongly coupled PMHS this means detection of ac magnetic fields with frequency of order $2g_{21}$, which in the present case is about 200 MHz.
 Since $\omega_1$ can be tuned via $B_0$,  the working frequency of the device can be changed, with the downside of reducing the sensitivity. Although a dedicated model has not yet been written, the reduction is expected to be $O(1)$ for a tuning of about 100\,MHz \cite{Crescini2020}, where the dispersive coupling might even increase the sensitivity as $\alpha\rightarrow1$.
We refer to this frequency range as the dynamical bandwidth of the magnetometer \cite{Crescini2020}.
More in general, the operational frequency of this class of magnetometers is limited from above by the larger strong coupling which can be reached, and from below by a combination of pump noise and filtering. While the former only depends on the hybrid system under consideration and can be of several GHz, the latter is mostly instrumental. To preserve the validity of Eq.\,(\ref{eq:sigmab1}), the $\mathrm{RAM}(\omega_b)$ must be reduced of $k_1$ to have $k_1\mathrm{RAM}(\omega_b) \lesssim k_B T_s$, which is increasingly difficult at lower $\omega_b$ due to the filter slope and to the higher pump noise. With present means, a lower limit for this frequency can be set around 10\,MHz. Arbitrarily low frequencies can be reached at the price of a much lower sensitivity.

	\begin{figure}%[h!]
	\includegraphics[width=.45\textwidth]{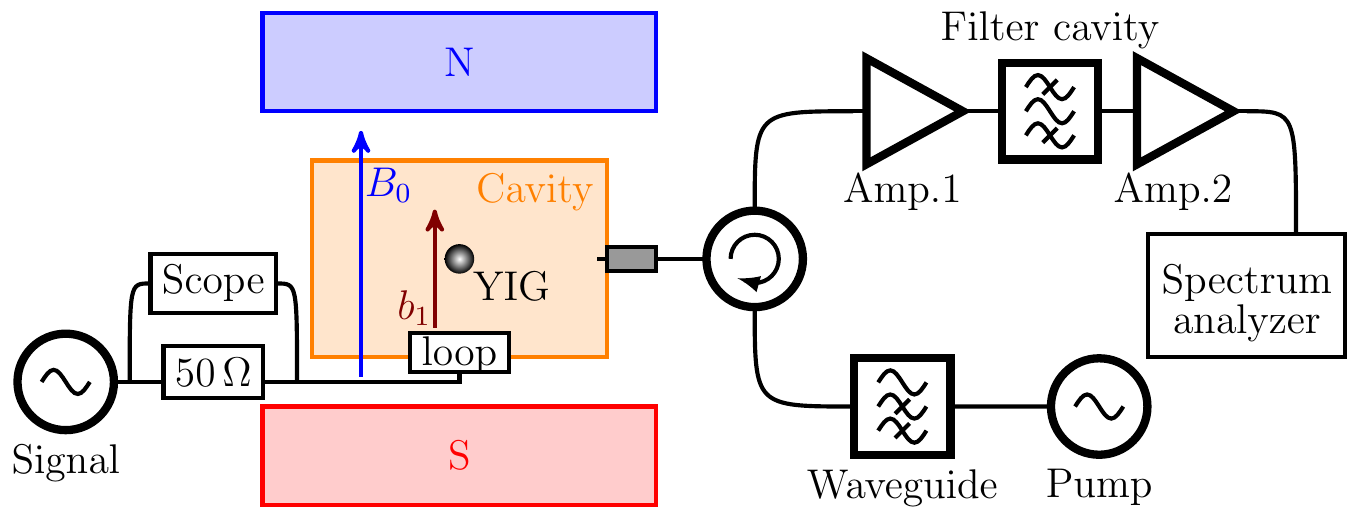}
	\caption{Scheme of the magnetometer demonstrator. The calibration signal $b_1$ (dark red) is provided by a generator connected to a loop on the holed side of the cavity, and is monitored with an oscilloscope. The cavity (orange) houses a YIG sphere (black), which is biased with the static field $B_0$ supplied by an electromagnet represented with its north (N, blue) and south (S, red) poles. The rf cavity field is extracted by an antenna with variable coupling (grey), and passes through a circulator before being amplified by Amp.1 and Amp.2, while the filter cavity avoids the saturation of the second amplifier. The microwave pumping line delivers to the cavity a monochromatic tone filtered with a waveguide. }
	\label{fig:exp_scheme}
	\end{figure}

\section{Experimental demonstration}
\label{sec:experimental}
To demonstrate the operation of the magnetometer we built the apparatus of  Fig.\,\ref{fig:exp_scheme}. It features a microwave copper cavity $32.1\,\mathrm{mm}$ long, with a quasi-rectangular section of  $14.8 \times 24.8\,\mathrm{mm}^2$, as the shorter sides were rounded. 
A 2\,mm-diameter YIG sphere is placed at the center of the cavity, where the rf magnetic field of the chosen $\mathrm{TM}_{102}$ mode is maximum. The mode frequency is $\omega_0/2\pi\simeq11.43$\,GHz, and the static field necessary to create the PMHS is $B_0 = \omega_0/\gamma \simeq 0.4\,$T, provided by an  electromagnet with a bore of 10\,cm diameter. The resulting hybrid modes resonate at $\omega_1\simeq (2\pi)\,11.33\,$GHz and $\omega_2\simeq (2\pi)\,11.55$\,GHz, obtaining $2g_{21}=\omega_2-\omega_1\simeq (2\pi)\,220\,$MHz close to the full hybridization, and thus with $\alpha\simeq0.5$.

The cavity is equipped with a hole on one side, hosting a $\sim1\,\mathrm{cm}$-diameter loop to generate a known test field $b_1$ parallel to the static field $B_0$. 
A scope measures the current provided to the loop to infer the generated field.
The current-to-field conversion constant was measured by analysing the effect of the loop dc-field on the Larmor frequency of the YIG sample, accessible by working in the dispersive regime.
The conversion constant results $4.8\,\mu\mathrm{T/A}$, and is compatible with the calculated loop field. Two more measurement followed, which were performed using a sensing coil with a diameter of about 1\,mm, much smaller than the loop.
In the first one, a $\sim100\,\mathrm{Hz}$-frequency field amplitude is compared with one at high frequency.
We found a 10\% decrease of the field amplitude due to the operation at 220\,MHz, which is probably related the loop impedance. The second one is performed to take into account the screening effect of the cavity walls. We used the sensing coil to measure the loop field at the position of the YIG sphere with and without the cavity. The ratio of the two gives a reduction factor of $k_b = 0.09 \pm 0.03$. The overall current-to-field conversion results $4.8\,\mu\mathrm{T/A} \times 0.9 \times k_b = \left( 0.39 \pm 0.11 \right) \mu\mathrm{T/A}$.
The  generator for the pump $A_p$ could deliver a power up to 26 dBm, injected to a in-house fabricated waveguide with a cut off frequency at about $\omega_0$. The waveguide has a cross section of $12.8\times 6.3$\,mm$^2$. The antenna is movable in order to choose appropriate coupling.
The two amplifiers are  low noise high-electron-mobility transistor (HEMT), with nominal gains $G_1= 34$ dB, $G_2=35$ dB, and the band-pass filter is done with a resonant cavity tuned to $\omega_2$. The readout is done with a spectrum analyzer.
%to avoid saturation of the second HEMT.

\begin{figure}%[h!]
	\includegraphics[width=.5\textwidth]{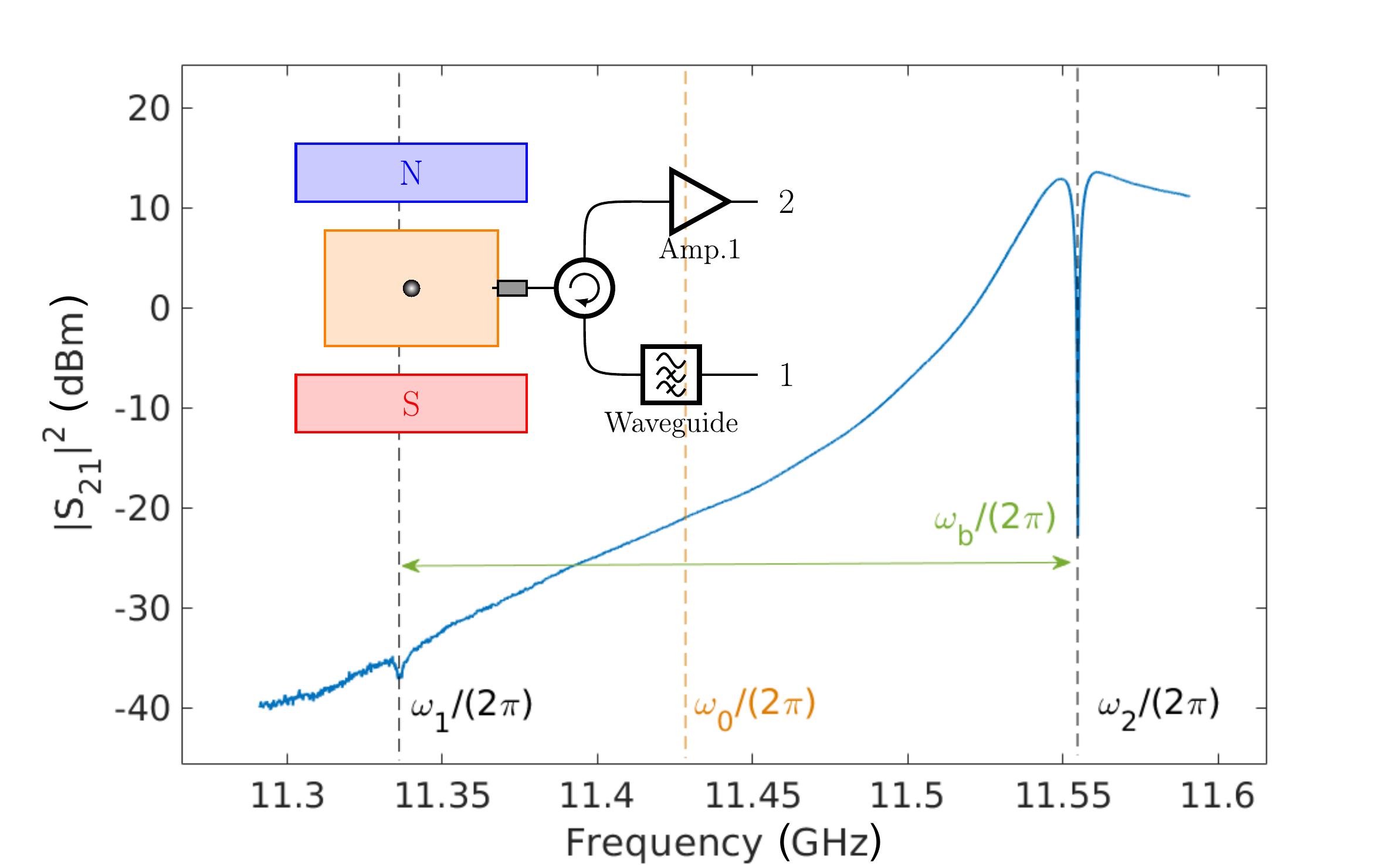}
	\caption{$S_{21}$ spectrum of the system reported as inset. The two absorption profiles are at the frequencies of the two hybrid modes $\omega_1$ and $\omega_2$. The waveguide cut-off is roughly at the frequency $\omega_0$. 
	%This is the experimental realization of the scheme of Fig.\,\ref{fig:scheme}. 
	The setup in the inset has the same colour legend described in Fig.\,\ref{fig:exp_scheme}.}
	\label{fig:s21}
	\end{figure}
	
Fig.\,\ref{fig:s21} shows a spectroscopic measurement of the PMHS, from which we measure $\gamma_1\simeq (2\pi)\,2.2\,$MHz, $\gamma_2\simeq (2\pi)\,4.2\,$MHz.
Magnonic dissipation are limiting the coherence of this PMHS, as the magnon linewidth is estimated to be of about 3.6\,MHz. The asymmetry of the modes was adjusted to match the waveguide transmission, and is not expected to have an important impact on the sensor.
The antenna couplings to the hybrid modes are $\beta_1\simeq0.3$ at $\omega_1$, and $\beta_2\simeq 1$ at $\omega_2$. 
Even with the band-pass filter, these couplings are necessary to avoid the saturation of the amplifier due to the pump, and still extract some power from $\omega_1$. 
The reflected tone power at $\omega_2$ is  reduced of more than 30\,dB. By injecting calibrated rf signals in the antenna input of the circulator (see Fig.\,\ref{fig:exp_scheme}), we measured $T_s\simeq 337\,$K, compatible with the amplifiers specifications. 
We fed the system with a pump of  power $A_p^2=0.2$\,mW at the antenna input.
%A stronger tone produces overheating of this setup, which in particular compromise the critical coupling of the antenna at $\omega_2$ by shifting the resonant frequency. This is a crucial problem for this test apparatus, as a large reflected microwave tone saturates the amplifier. The problem can be mitigated in a future apparatus by different means, e.\,g. by employing dedicated microwave filtering at the amplifier input, or including heat sinks.
Following Eq.\,(\ref{sens}), the expected sensitivity results
\begin{equation}
\sigma_{b_1} = 1.9\, {\rm pT}/\sqrt{\rm Hz} 
\label{expsens}.
\end{equation}
We remind that this spin-magnetometer is sensitive to a field at the frequency $2g_{21}\simeq (2\pi)\,220\,$MHz on a bandwidth as large as the linewidth $\gamma_1$. 

\begin{figure}
	\includegraphics[width=.5\textwidth]{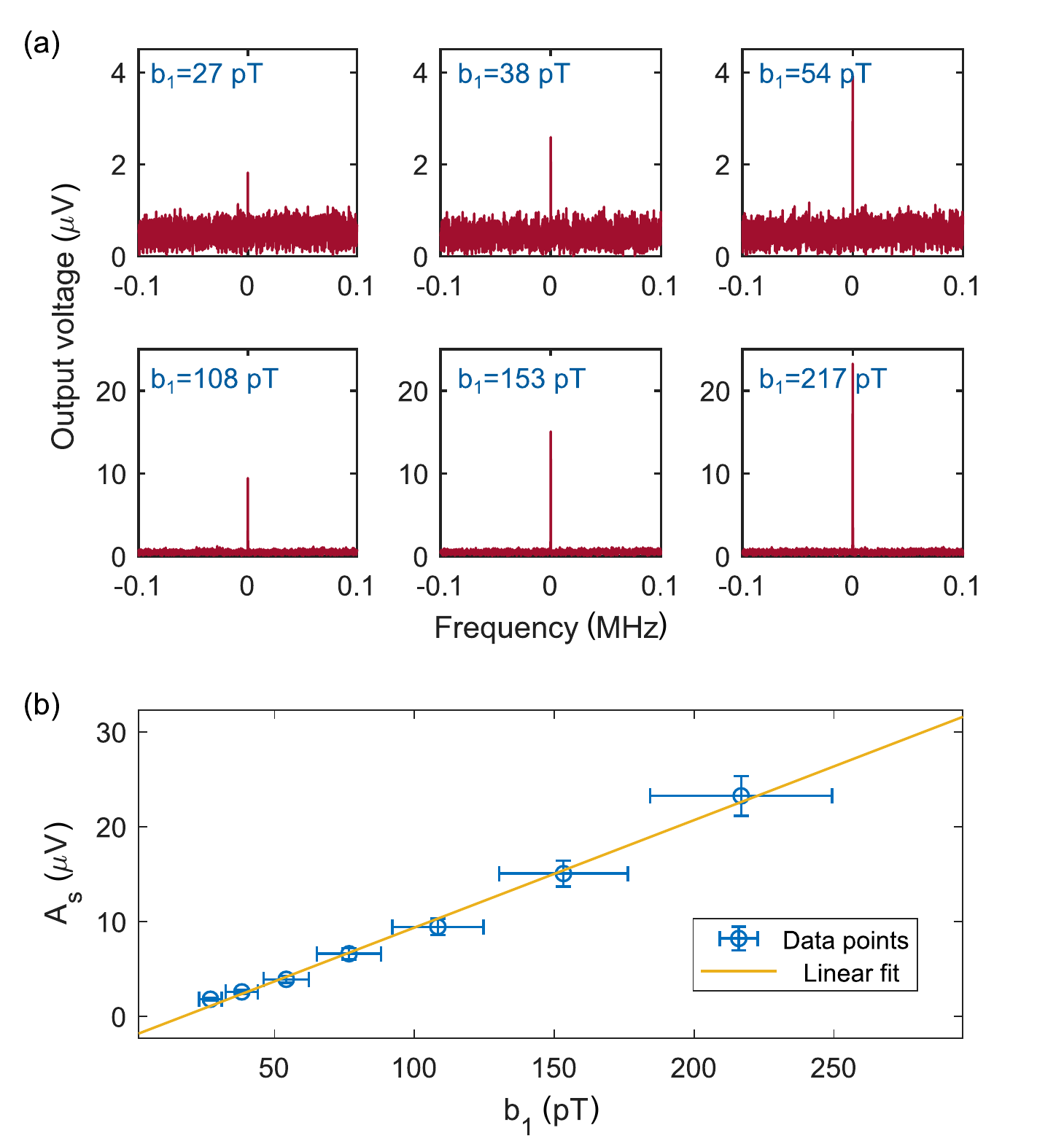}
	\caption{(a) Signal spectra obtained with different test loop fields. The $x$-axis is the frequency offset from $\omega_1/2\pi$, while the $y$-axis is the voltage read at the spectrum analyser. The top-left  spectrum corresponds to a $\mathrm{SNR}=1$ field with a resolution bandwidth of 100\,Hz, i.e. an integration time of 10\,ms. (b) Test measurement of the spin-magnetometer, showing the voltage at $\omega_1$ vs the calculated input field generated with the loop. The noise level was subtracted from the data points, which were then fitted to a linear function, giving a slope $m_\mathrm{fit}= (1.1\pm 0.3) \times 10^{5}\,{\rm V/T}$ and an intercept  $q_\mathrm{fit}= (-1.9 \pm 0.7)\,\mathrm{\mu V}$.
}
	\label{fig:calib}
	\end{figure}

The system was tested using the small loop placed on the cavity side hole (see Fig.\,\ref{fig:exp_scheme}). 
%The DC current-field proportionality was measured as $4.8\,\mu\mathrm{T/A}$, compatible with the calculated loop field. A 10\% decrease of this constant has been measured for operating it at 220 MHz. Moreover, to take into account the screening effect of the cavity walls, we used a sensing coil to measure the loop field at the distance of the YIG sphere with a) the loop in its standard position and  b) the loop in free space.  The ratio of the two gives a reduction factor $k_b=(0.09 \pm 0.03)$.
%By using current values ranging from 0.06\,mA to 0.5\,mA at 220\,MHz we obtained the experimental points shown in Figure \ref{fig:calib}. Reported amplitude values correspond to the voltage on a $R=50\,\Omega$ load at the signal frequency $\omega_1 \simeq (2\pi)\,11.33\,$GHz, to be compared with
Fig.\,\ref{fig:calib}a shows the spectra of the output signal for different input test fields. 
The experimental points reported in Fig.\,\ref{fig:calib}b shown the voltage $A_s$ measured on a $R=50\,\Omega$ load at the signal frequency $\omega_1 \simeq (2\pi)\,11.33\,$GHz vs the field generated with the loop. The errors on the values $A_s$ are estimated from repeated measurements with the same input field, they account for an  $\sim 18\%$ of the average value and are essentially due to instabilities in the experimental set-up. The uncertainty on the  input field, 28\%, is almost totally due to the determination of the loop current-to-field coefficient. Data have been fitted with a linear function, giving the slope $m_\mathrm{fit} = (1.1\pm 0.3) \times 10^{5}\,{\rm V/T}.$ The expected value, derived from Eq.\,(\ref{totpower}), is
% Voltage values will be compared to
%\begin{eqnarray}
%A_s (\omega_1)&=&\sqrt{R P_s(\omega_1) }\\
%& = & A_p \sqrt{R G_1 G_2 c_1  } \sqrt{\frac{\beta_2 \beta_1}{1+\beta_1}}\frac{\pi Q_2 \alpha }{2B_0 }b_1 \nonumber.
%\end{eqnarray}
\begin{align}
\begin{split}
\frac{\sqrt{R P_s(\omega_1) }}{b_1}& =
\sqrt{R G_1 G_2 c_1  } \frac{\pi \alpha Q_2^L A_p}{2\eta B_0}\\
& =(1.6 \pm 0.3) \times10^{5}\,\, {\rm V/T},
\end{split}
\end{align}
%In the figure the errors on the amplitude are estimated from multiple measurements with the same input field, they account about an  $ 18\%$ of the average value and are essentially due to instabilities in the experimental set-up. 
%The uncertainty on the  input field is almost totally due to the correction factor $k_b$. 
%The slope resulting from the plot is
%\begin{equation}
%m_\mathrm{fit} = (1.1\pm 0.3) \times 10^{5}\,{\rm V/T},
%\label{coeff}
%$\end{equation}
%the expected value is $(1.6 \pm 0.3) \times10^{5}$ V/T, 
where the total gain $G_1 G_2 c_1 = 56\pm1$ dB has been measured independently.
The discrepancy can be due to unaccounted losses in the cables, but essentially the measured value agrees with the expected one, confirming the validity of the proposed device.

From this data we obtained the noise $P_n= - 99.8$ dBm with a resolution bandwidth of ${\rm RBW}=100$ Hz, resulting in a voltage sensitivity of $2.2 \times 10^{-7} $ V/$\sqrt{\rm Hz}$. By using the measured transduction coefficient $m_\mathrm{fit}$, the field sensitivity results
\begin{equation}
\sigma_{b_1}^s = (2.0 \pm 0.7)\, {\rm pT}/\sqrt{\rm Hz},
\end{equation}
which agrees with the calculated one given by Eq.\,(\ref{expsens}).
% and shows that even a very simple device, not yet optimized, can be used as a very performing magnetometer.
The noise level at the cavity output results $P_n^i = P_n /(G_1 G_2 c_1) = (2.6\pm 0.5) \cdot 10^{-21}$ W/Hz.
In the absence of waveguide filtering, the pump noise  at $\omega_1$ would be $\sim10^{-19}\,\mathrm{W/Hz}$, orders of magnitude higher than the measured thermal noise. By removing the waveguide we verified that the noise grows, and test signals has to be increased to obtain the same SNR. We conclude that the pump noise has been drastically reduced in this setup, and that the limiting background of the apparatus is mostly due to room temperature thermodynamic fluctuations. 
The negative offset's origin is unclear, but it might be attributed to systematic uncertainties in the measurement. 
%As described previously, the uncertainty on the input field amplitude is large 

The time response of the system is limited by the relaxation time of the hybrid modes and is of order 100\,ns.
The maximum measurable field is given by the condition $Q_1b_1\ll B_0$, which holds up to tens of $\mu$T amplitudes.
The operational bandwidth  is an interval of width $\gamma_1$ centered at $\omega_1$. The CMPs frequencies are tuned by varying the static field $B_0$, allowing us to obtain an effective bandwidth  of hundreds of MHz \cite{Crescini2020}. 
The loop on the cavity side, which has been used for testing the setup, can be coupled to an external pick-up coil. With such a scheme it will be possible to measure magnetic fields not directly located on the YIG sample. By properly designing the loop impedance, optimum matching conditions can be achieved with the pick-up loop.
We expect also the factor $k_b$ to be much closer to one in an optimized apparatus, allowing for a good transduction of an external signal to the PMHS.

\section{Outlook and conclusions}
\label{sec:outlook}
The sensitivity can be further pushed by increasing the tone power, however, this value is limited due either to non-linearities of the YIG or thermal-induced instabilities that may arise. By using a cryogenically  cooled device coupled with an ultra low noise first-stage amplifier, the limiting noise could be reduced to the standard quantum limit.  By replacing the factor $k_B T_s$ with $\hbar \omega$ in Eq.\,(\ref{sens}), the quoted value $\sigma_{b_1} \simeq 8 $ fT/$\sqrt{\rm Hz}$ can be further reduced to  $\sigma_{b_1} \simeq 0.3 $ fT/$\sqrt{\rm Hz}$.

%Depending on the envisioned application, a number of different configurations can be designed for this instrument. 
As the device sensitivity is not dependent on its size, one can engineer a much smaller PMHS, consisting for example in printed circuits or dielectric cavities coupled to a suitable magnetic material. Miniaturization may allow for simpler construction and implementation of the device. Multiplexing is also a possibility, for example by means of superconducting circuits \cite{doi:10.1063/1.1791733,doi:10.1063/1.2803852,doi:10.1063/1.4973872} resonating within the magnetometer bandwidth, which have demonstrated quality factors up to $10^8$ \cite{1439774,PhysRevB.94.014506,doi:10.1063/1.4919761}.
Thanks to the present demonstration of the measurement scheme, forthcoming works will study alternative implementations of the magnetometer, and the device itself on a more detailed level, including for example the experimental demonstration of its dynamical bandwidth, the long-term stability of the setup and an improved sensitivity.

In conclusion, we devised and tested a spin-magnetometer based on the phase modulation of cavity magnon polaritons. Our prototype demonstrated the working principle of the device by showing a pT field sensitivity, and an optimized version of the same setup could measure magnetic fields with $\mathrm{fT/\sqrt{Hz}}$ precision. The tested design is not the only realization of the experimental scheme, which can be implemented in a number of different configurations. The magnetometer has a variable input impedance, a large dynamic range, and a remarkable sensitivity, allowing for its application for example in the field on magnetic imaging.
All these features are unprecedented, and make this instrument the first of a class of magnetometers based on photon-magnon hybrid system.

\section*{Acknowledgments}
We acknowledge Enrico Berto, Mario Tessaro and Fulvio Calaon for their work on the building and operation of the prototype.
We thank Federico Chiossi, Caterina Braggio, Antonello Ortolan and Pierre Sikivie for the stimulating discussion regarding the physics of the instrument.
The support of INFN-Laboratori Nazionali di Legnaro and of INFN-CNTT is deeply acknowledged.
The authors declare that there is a pending patent application related to this research.

\bibliographystyle{apssamp}
%\bibliography{hybridSpinMag}

\end{document}